\documentclass[%
 reprint,
 amsmath,amssymb,
 aps,
 prl,
 longbibliography,
 lengthcheck,%
]{revtex4-1}

\usepackage{graphicx}
\usepackage{dcolumn}
\usepackage{bm}
\usepackage{hyperref}

\begin{document}

\preprint{APS/123-QED}

\title{Non-integral form of the Gross-Pitaevskii equation for polarized molecules
}

\author{Pavel A. Andreev}%
\email{andreevpa@physics.msu.ru}
 \affiliation{Department of General Physics, Physics Faculty, Moscow State
University, Moscow, Russian Federation.}





\date{\today}

\begin{abstract}
The Gross-Pitaevskii equation for polarized molecules is an integro-differential equation, consequently it is complicated for solving. We find a possibility to represent it as a non-integral nonlinear Schrodinger equation, but this equation should be coupled with two linear equations describing electric field. These two equations are the Maxwell equations. We recapture the dispersion of collective excitations in the three dimensional electrically polarized BEC with no evolution of the electric dipole moment directions. We trace the contribution of the electric dipole moment. We explicitly consider the contribution of the electric dipole moment in the interaction constant for the short-range interaction. We show that the spectrum of dipolar BEC reveals no instability at repulsive short-range interaction. Nonlinear excitations are also considered. We present dependence of the bright soliton characteristics on the electric dipole moment.

\end{abstract}

\maketitle


\section{Introduction}

Basic properties of collective excitations in a dipolar Bose-Einstein condensate (BEC) were studied about ten years ago, generalization of the Gross-Pitaevskii equation for polarized particles were suggested and used \cite{Yi PRA 00}-\cite{Giovanazzi EPJD 04}. Since when, evolution of collective excitations in the dipolar BEC with dipoles being strictly parallel to an external field has
attracted a lot of attention. Rotons and maxons are most
fascinating features of the polar BEC spectrum revealed by a dilute two-dimensional Bose-condensed dipolar gas \cite{Santos PRL 03}, \cite{Ronen PRL 07}, \cite{Jona-Lasinio 13}. Instability of the
three-dimensional dipolar BEC leads to shift of the attention on the quasi two- and one-dimensional BEC \cite{Fischer PRA 06R}. Dispersion of excitations and other properties of ultracold Fermi and Bose molecules has been studied \cite{Lahaye RPP 09}-\cite{Sieberer ar 11}.

In our previous papers we have considered the spectrum of polarized BEC including evolution of electric dipole moment direction \cite{Andreev arxiv 12 02}-\cite{Andreev TransDipBEC12} using the method of quantum hydrodynamics (QHD) developed in Ref.s \cite{Andreev PRB 11}-\cite{Andreev IJMP B 13}. In this paper we consider simpler case. It is usually considered at studying of polarized ultracold gases, when all dipoles are parallel to an external electric field \cite{Lahaye RPP 09}, \cite{Baranov CR 12}.

Derivation of the QHD equations for electrically polarized BEC including evolution of the electric dipole direction was described briefly in appendixes in Ref.s \cite{Andreev arxiv 12 02} and \cite{Andreev arxiv Pol 11}. Full QHD scheme for polarized particles without the short-range interaction was described in Ref. \cite{Andreev PRB 11}. In this paper we describe the derivation of the continuity and Euler equations for the fully polarized electric dipole moments, which are used here to consider the dispersion of collective excitations. Our analysis allows to represent the integro-differential Gross-Pitaevskii equation for polarized particles in a non-integral form.

One of the most important points we consider in the paper is the explicit form of the potential energy of electric dipole interaction.  Electric dipoles create the electric field and the electric field should satisfy to the fundamental Maxwell equations. Thus, we present explicit form of the potential energy and trace contribution all terms of the potential energy in the dispersion of excitations.
We present the dispersion of electrically dipolar BEC in comparison with nonpolar
BEC reveals the Bogoliubov spectrum, characterized by a linear (phonon) dispersion at low momenta,
and a quadratic dispersion at large momenta (which gives the single-particle dispersion at neglecting the third order interaction by the interaction radius term \cite{Andreev PRA08}). We find that the dispersion of electrically polarized three-dimensional Bose-Einstein condensate shows no instability.

This paper is organized as follows. In Sec. II we describe a representation of the integral Gross-Pitaevskii equation in a non-integral form and present corresponding hydrodynamic equations. In Sec. III we present the derivation of the quantum hydrodynamic equations from first principles in the approximation corresponding to the Gross-Pitaevskii equation. In Sec. IV  we consider the excitation spectrum of three-dimensional plane waves. In Sec. V we describe the bright soliton propagation in electrically polarized BEC in two cases: soliton propagation parallel and perpendicular to an external electric field. In Sec. VI brief conclusions are presented.

\section{Representation of the GP equation as a non-integral non-linear Schrodinger equation}

In this paper we are going to discuss different form of presentations of the GP equation for electrically polarized molecules. It, for instance, can be written as
$$\imath\hbar\partial_{t}\Phi(\textbf{r},t)=\Biggl(-\frac{\hbar^{2}}{2m}\triangle+g\mid\Phi(\textbf{r},t)\mid^{2}$$
\begin{equation}\label{NI GP nlse int polariz some frame}-d^{\beta}d^{\gamma}\int d\textbf{r}' G^{\beta\gamma}(\textbf{r},\textbf{r}')n(\textbf{r}',t)\Biggr)\Phi(\textbf{r},t),\end{equation}
where $m$ is the mass of molecule, $\hbar$ is the reduced Planck constant, $\textbf{d}$ is the electric dipole moment of molecule, and $n(\textbf{r},t)=\Phi^{*}(\textbf{r},t)\Phi(\textbf{r},t)$ is the concentration of particles. The second term in the right-hand side of equation (\ref{NI GP nlse int polariz some frame}) presents the short-range interaction, and $g$ is the interaction constant defined via integral of a short-range interaction potential.
The last term in this equation describes dipole-dipole interaction and contains energy of dipole-dipole interaction $U_{dd}$ written via the Green function $G^{\alpha\beta}$. One can see that the GP equation (\ref{NI GP nlse int polariz some frame}) is a nonlinear integral equation, and the integral term contains a nonlinearity. This equation can be presented as a set of two hydrodynamic equations \cite{Lahaye RPP 09}, the continuity equation and the Euler equation. Nevertheless, corresponding Euler equation also contains the nonlinear integral term (see for example Ref. \cite{Lahaye RPP 09} formula (4.9)). Thus, equation (\ref{NI GP nlse int polariz some frame}) is very complicated. Below, in this section, we will show that this equation can be represented in nonintegral form. A nonlinear term appears instead of the integral term, and we come to the set of equations instead of one equation (\ref{NI GP nlse int polariz some frame}). Fortunately, these two additional equations are linear equations. Solving of a nonintegral equation coupled with two linear equations is easier than solving of an integral nonlinear equation.

Potential of the electric field caused by an electric dipole appears as $\varphi=-(\textbf{d}\nabla)(1/r)$ \cite{Landau 2}, consequently electric field has the form of $\textbf{E}=-\nabla\varphi$$=(\textbf{d}\nabla)\nabla(1/r)$ or it can be rewritten in the tensor form $E^{\alpha}=d^{\beta}\nabla^{\beta}\nabla^{\alpha}(1/r)$. Energy of interaction of two dipoles comes as $U_{dd}=-\textbf{d}_{2}\textbf{E}_{21}$, where $\textbf{d}_{2}$ is the electric dipole moment of the second dipole, and $\textbf{E}_{21}$ is the electric field caused by the first dipole acting on the second dipole. The explicit form of dipole-dipole interaction energy is $U_{dd}=-d^{\alpha}_{2}d^{\beta}_{1}\nabla^{\beta}\nabla^{\alpha}(1/r)$. It can be rewritten via the notion of Green function $U_{dd}=-d^{\alpha}_{2}d^{\beta}_{1}G^{\alpha\beta}$, where $G^{\alpha\beta}=\partial^{\alpha}\partial^{\beta}\frac{1}{r}$ is the Green function of dipole-dipole interaction.
Using well-known identity
\begin{equation}\label{NI GP togdestvo}-\partial^{\alpha}\partial^{\beta}\frac{1}{r}= \frac{\delta^{\alpha\beta}-3r^{\alpha}r^{\beta}/r^{2}}{r^{3}}+\frac{4\pi}{3}\delta^{\alpha\beta}\delta(\textbf{r}),\end{equation}
we can represent energy of dipole-dipole interaction in the explicit form. This form contains the term proportional to the Dirac delta function. At mechanical description of interaction of two dipoles we can put this term away. However, it is very important to keep this term in the theory of material fields, such as quantum hydrodynamics and the Gross-Pitaevskii equation appearing as a part of the quantum hydrodynamic description \cite{Andreev PRA08}. This notice becomes especially important because we integrate potential of dipole-dipole interaction in equation (\ref{NI GP nlse int polariz some frame}) over whole space including point $\textbf{r}=\textbf{r}'$. We get that electric field does not satisfy to the Maxwell equations in the absence of a term proportional to the Dirac delta function. Choosing a frame of reference with z axes parallel to $\textbf{d}$ and neglecting the delta function term we come to potential energy $U_{dd,z}=-d^{2} (1-3\cos^{2}\vartheta)/r^{3}$. It was used for generalization of the GP equation on the system of Bose particles having electric dipole moment suggested in Ref.s \cite{Yi PRA 00}, \cite{Goral PRA 00}, \cite{Santos PRL 00}, it is also discussed in review \cite{Lahaye RPP 09}. This equation does not contain whole electric field that leads to mistreating of dipole-dipole interaction. The delta function term in dipole-dipole interaction may be included in the short range interaction,
which is also proportional to the Dirac delta function in the pseudopotential description \cite{Landau 9}. Explicit consideration of including of the delta-functional term in the short-range interaction is presented in Appendix. This consideration shows that extracting of dipole contribution in the short range interaction confirm our results based on consideration of full energy of dipole-dipole interaction. Moreover, we are going to show that the consideration of full energy of dipole-dipole interaction gives some benefits. One of them has been mentioned already. We have deal with the electric field satisfying to the Maxwell equations. Explicitly introducing this electric field we can rewrite integro-differential equations (the Gross-Pitaevskii equation or corresponding equations of quantum hydrodynamics) in non-integral form. We also include the Maxwell equations in the considering set of equations. Despite the fact that numerical methods of solving of the integral Gross-Pitaevskii equation for dipolar BEC have been developed, we can admit that solving of the non-integral Gross-Pitaevskii equation and Maxwell equations is much easier and allows to get analytical results. As example, we present an exact analytical solution for the bright soliton in electrically polarized BEC. Moreover, in this paper the interaction constant of the short-range interaction $g$ does not depend on molecule dipole value.

We can include the energy of dipoles in an external electric field $-\textbf{d}\textbf{E}_{ext}$ in the GP equation (\ref{NI GP nlse int polariz some frame}).
Electric dipoles create the electric field, including their internal field, we can rewrite the last term in equation (\ref{NI GP nlse int polariz some frame}) in form of $-\textbf{d}\textbf{E}$ introducing the electric field caused by electric dipoles. In result we find
\begin{equation}\label{NI GP nlse int polariz some frame non Int}\imath\hbar\partial_{t}\Phi(\textbf{r},t)=\Biggl(-\frac{\hbar^{2}}{2m}\triangle+g\mid\Phi(\textbf{r},t)\mid^{2}-\textbf{d}\textbf{E}\Biggr)\Phi(\textbf{r},t),\end{equation}
where $\textbf{E}=\textbf{E}_{ext}+\textbf{E}_{int}$, $\textbf{E}_{int}$ is the electric field caused by dipoles. However introducing additional physical variable we have to present equation for this variable. Looking on the explicit form of internal electric field
$$E^{\alpha}_{int}=d^{\beta}\int d\textbf{r}' G^{\alpha\beta}(\textbf{r},\textbf{r}')n(\textbf{r}',t)$$
we can obtain that this electric field satisfies to the Maxwell equations
$$\nabla\textbf{E}(\textbf{r},t)=-4\pi \nabla\textbf{P}(\textbf{r},t)$$
\begin{equation}\label{NI GP field good div}=-4\pi (\textbf{d}\nabla) n(\textbf{r},t)=-4\pi (\textbf{d}\nabla) (\Phi^{*}(\textbf{r},t)\Phi(\textbf{r},t)),\end{equation}
and
\begin{equation}\label{NI GP field good curl}\nabla\times\textbf{E}(\textbf{r},t)=0.\end{equation}
Equation (\ref{NI GP field good div}) is also called the Poisson equation with the charge equal to zero and explicit writing of the electric displacement field via the electric field $\textbf{E}$ and the polarization $\textbf{P}$. Equation (\ref{NI GP field good curl}) is the quasi static form of the  Faraday's law of induction.
For getting Maxwell equations we have to use the full Green function of dipole-dipole interaction.
Equation (\ref{NI GP nlse int polariz some frame non Int}) is the non-integral form of GP equation for electrically polarized particles.

Non-linear Schrodinger equation (\ref{NI GP nlse int polariz some frame non Int}) can be represented as a set of two hydrodynamic equation, they are the continuity equation
\begin{equation}\label{NI GP cont eq from GP}
\partial_{t}n+\nabla(n\textbf{v})=0,\end{equation}
and the Euler equation
$$mn(\partial_{t}+\textbf{v}\nabla)v^{\alpha}-\frac{\hbar^{2}}{4m}\partial^{\alpha}\triangle n
+\frac{\hbar^{2}}{4m}\partial^{\beta}\biggl(\frac{1}{n}(\partial^{\alpha}n)(\partial^{\beta}n)\biggr)$$
\begin{equation}\label{NI GP bal imp eq short from GP}=-gn\partial^{\alpha}n+
nd\partial^{\alpha}(\textbf{l}\textbf{E}),\end{equation}
where $\textbf{l}$ is the unit vector in the direction of the polarization formed by the external field, $n$ is the particle concentration, $n=\Phi^{*}\Phi$, and $\textbf{v}$ is the velocity field.

In the last term of the equation (\ref{NI GP bal imp eq short from GP}), $nd\textbf{l}$ is the polarization of medium for fully polarized dipoles with no evolution of dipole direction. This term appears instead of
$P^{\beta}(\textbf{r},t)\partial^{\alpha}E^{\beta}(\textbf{r},t)$, where $\textbf{P}$ is the polarization of medium for general case including the dipole direction evolution around the equilibrium position formed by the external electric field. Below we present the derivation of equation (\ref{NI GP bal imp eq short from GP}) from a microscopic theory (from the many particle Schrodinger equation).

\section{Derivation of QHD equations from Schrodinger equation}

We have discussed the Gross-Pitaevskii equation for polar molecules and corresponding equations of hydrodynamics as a master equations. In this section the derivation of the quantum hydrodynamic equations from the first principles is described.

Starting from the many-particle Schrodinger equation
$$\imath\hbar\partial_{t}\psi(R,t)=\hat{H}\psi(R,t),$$
where $R=(\textbf{r}_{1}, ..., \textbf{r}_{N})$ is the set of coordinate of $N$ particles, with the Hamiltonian
$$\hat{H}=\sum_{i}\Biggl(\frac{1}{2m_{i}}\textbf{p}_{i}^{2}-d_{i}^{\alpha}E_{i,ext}^{\alpha}\Biggr)$$
\begin{equation}\label{NI GP Hamiltonian}+\frac{1}{2}\sum_{i,j\neq i}U_{ij}-\frac{1}{2}\sum_{i,j\neq i}\Biggl(d_{i}^{\alpha}d_{j}^{\beta}G_{ij}^{\alpha\beta}\Biggr),\end{equation}
we construct a system of QHD equations for particles having an electric dipole moment $d_{i}^{\alpha}$. The following
designations are used in the Hamiltonian (\ref{NI GP Hamiltonian}):
$p_{i}^{\alpha}=-\imath\hbar\partial_{i}^{\alpha}$, $E_{i,ext}^{\alpha}$ is the
electric field, $m_{i}$ is mass of particles, $\hbar$ is the reduced Planck constant, and
$G_{ij}^{\alpha\beta}=\partial_{i}^{\alpha}\partial_{i}^{\beta}1/r_{ij}$
is the Green function of dipole-dipole
interaction, $U_{ij}=U(|\textbf{r}_{i}-\textbf{r}_{j}|)$ is the potential of the short-range interaction. Writing Hamiltonian (\ref{NI GP Hamiltonian}) we suppose that particles have electric dipole moment, but directions of electric dipole moments are, in general case, different, but they can be polarized in the direction of external field. The dipole-dipole interaction leads to the evolution of both positions of particles and directions of the electric dipole moments.

The first step in the construction of the QHD method is to determine the concentration of particles in the vicinity of $\textbf{r}$ in the physical space. If we define the concentration of particles as quantum average of the concentration operator in the coordinate representation $\hat{n}=\sum_{i}\delta(\textbf{r}-\textbf{r}_{i})$ we obtain
\begin{equation}\label{NI GP def density}n(\textbf{r},t)=\int dR\sum_{i}\delta(\textbf{r}-\textbf{r}_{i})\psi^{*}(R,t)\psi(R,t),\end{equation}
where $dR=\prod_{p=1}^{N}d\textbf{r}_{p}$.

Differentiation of $n(\textbf{r},t)$ with respect to time and applying of the Schr\"{o}dinger equation with Hamiltonian (\ref{NI GP Hamiltonian}) leads to the
continuity equation
$\partial_{t}n+\nabla\textbf{j}=0,$
where the current density takes the form of
\begin{equation}\label{NI GP def of current of density}\textbf{j}(\textbf{r},t)=\int dR\sum_{i}\delta(\textbf{r}-\textbf{r}_{i})\frac{1}{2m_{i}}\biggl(\psi^{*}(R,t)(\textbf{p}_{i}\psi(R,t))+h.c.\biggr).\end{equation}
In hydrodynamics one usually has deal with velocity field $\textbf{v}(\textbf{r},t)$ instead of the current density $\textbf{j}(\textbf{r},t)$, thus we have to present a connection of $\textbf{j}$ and $\textbf{v}$.

The velocity of i-th particle $\textbf{v}_{i}(R,t)$ is determined by equation
\begin{equation}\label{NI GP vel of i part}\textbf{v}_{i}(R,t)=\frac{1}{m_{i}}\nabla_{i}S(R,t),\end{equation}
where $S(R,t)$ presents the phase of the wave function
$$\psi(R,t)=a(R,t) \exp\biggl(\frac{\imath S(R,t)}{\hbar}\biggr).$$
The quantity $\textbf{v}_{i}(R,t)$ describes the current of probability connected with the motion of $i$-th particle, in general case $\textbf{v}_{i}(R,t)$ depends on coordinates of all particles of the system $R$, where $R$ is the totality of 3N coordinates of N particles of the system $R=(\textbf{r}_{1}, ..., \textbf{r}_{N})$. Velocity field $\textbf{v}(\textbf{r},t)$ is the velocity of the
local centre of mass and determined by equation $\textbf{j}(\textbf{r},t)=n(\textbf{r},t)\textbf{v}(\textbf{r},t)$.
This means that $\textbf{u}_{i}(\textbf{r},R,t)=\textbf{v}_{i}(R,t)-\textbf{v}(\textbf{r},t)$ is a quantum equivalent of the thermal velocity.

A momentum balance equation can be derived by differentiating current density (\ref{NI GP def of current of density}) with respect to time:
\begin{equation} \label{NI GP bal of imp gen}\partial_{t}j^{\alpha}(\textbf{r},t)+\frac{1}{m}\partial^{\beta}\Pi^{\alpha\beta}(\textbf{r},t)=\frac{1}{m}F^{\alpha}(\textbf{r},t),\end{equation}
where $F^{\alpha}(\textbf{r},t)$ is a force field and
$\Pi^{\alpha\beta}(\textbf{r},t)$ is the momentum current density tensor.

Performing explicit separation of particles' thermal movement with velocities $\textbf{u}_{i}(\textbf{r},R,t)$ and the collective movement of particles with velocity $\textbf{v}(\textbf{r},t)$ in equations of continuity and of the momentum balance (\ref{NI GP bal of imp gen}) we can find that the tensor $\Pi^{\alpha\beta}(\textbf{r},t)$ takes  form of
$$\Pi^{\alpha\beta}(\textbf{r},t)=mnv^{\alpha}v^{\beta}+p^{\alpha\beta}+T^{\alpha\beta}.$$
In this formula $p^{\alpha\beta}(\textbf{r},t)$ is the tensor of kinetic pressure. This tensor tends to zero by
letting $\textbf{u}_{i}\rightarrow 0$. The tensor
$T^{\alpha\beta}$ is proportional to $\hbar^{2}$ and has a purely
quantum origin. For the system of noninteracting particles,
this tensor is
\begin{equation}\label{NI GP Bom one part}T^{\alpha\beta}=-\frac{\hbar^{2}}{4m}\partial^{\alpha}\partial^{\beta}n+\frac{\hbar^{2}}{4m}\frac{1}{n}(\partial^{\alpha}n)(\partial^{\beta}n)
.\end{equation} This term is called the quantum Bohm potential.

As the particles of the system under consideration interact via long-range forces the approximation of the self-consistent field is sufficient to analyze collective processes. With the use of this approximation two-particle functions appiaring in the momentum balance equation can be split into a product of single-particle functions. Taken in the approximation of self-consistent field, the set of QHD equation, continuity equation and momentum balance equation has the form
\begin{equation}\label{NI GP cont eq}
\partial_{t}n+\nabla(n\textbf{v})=0 ,\end{equation}
and
$$mn(\partial_{t}+\textbf{v}\nabla)v^{\alpha}+\partial_{\beta}T^{\alpha\beta}=F^{\alpha}_{SRI}
$$
\begin{equation}\label{NI GP bal imp eq} +P^{\beta}\partial^{\alpha}E_{ext}^{\beta}+P^{\beta}\partial^{\alpha}\int d\textbf{r}'G^{\beta\gamma}(\textbf{r},\textbf{r}')P^{\gamma}(\textbf{r}',t).\end{equation}
Let's discuss the physical meaning of terms on the right-hand side of (\ref{NI GP bal imp eq}). Force density of the short-range interaction $F^{\alpha}_{SRI}(\textbf{r},t)$ was considered in Ref. \cite{Andreev PRA08}. It was found that in the first order by the interaction radius the force density has form corresponding to the Gross-Pitaevskii approximation $F^{\alpha}_{SRI}(\textbf{r},t)=-g n\partial^{\alpha}n$. The second term is the effect of non-uniform external electric field on the polarization density. It should be noted that the form of this term is distinct from the expression that describes the force affecting a single dipole. The third term of the equation (\ref{NI GP bal imp eq}) describes a force field that represents interactions between particles, namely the  dipole-dipole interaction.

Note, that for a three-dimensional system of particles the momentum balance equation (\ref{NI GP bal imp eq}) may be written down in terms of electrical intensity of the field that is created by dipole moments of the particle system:
$P^{\beta}(\textbf{r},t)\partial^{\alpha}E^{\beta}(\textbf{r},t)$,
where $\textbf{E}_{int}(\textbf{r},t)$ satisfy to the Maxwell equations (\ref{NI GP field good div}) and (\ref{NI GP field good curl}).

The method we develop in this work is valid both for bosons and fermions. The type of statistics that particles are subject to affects the calculation of many-particle functions (correlations) that evolve in the momentum balance equation (\ref{NI GP bal imp eq}) and are neglected in the self-consistent field approximation. A method for the calculation of correlations in the QHD equations has been developed in papers ~\cite{MaksimovTMP 1999,MaksimovTMP 2001}.

The polarization evolves in the momentum balance equation (\ref{NI GP bal imp eq}) is
\begin{equation}\label{NI GP def polarization}P^{\alpha}(\textbf{r},t)=\int dR\sum_{i}\delta(\textbf{r}-\textbf{r}_{i})\psi^{*}(R,t)\hat{d}_{i}^{\alpha}\psi(R,t).\end{equation}
This function presents general definition of the polarization appearing in the quantum hydrodynamics. It can be replaced with $nd\textbf{l}$ at assumption that all dipoles are parallel to each other. If we need to consider the dipole direction evolution we can derive equations for the polarization $\textbf{P}$, and, therefore, expand the set of the quantum hydrodynamic equations (see for example \cite{Andreev arxiv 12 02}, \cite{Andreev LongetDipFermi12}, \cite{Andreev PRB 11}).

\section{Excitations}
\subsection{calculation of the excitation dispersion}

Solving the set of QHD equations (\ref{NI GP field good div}), (\ref{NI GP field good curl}), (\ref{NI GP cont eq from GP}), and (\ref{NI GP bal imp eq short from GP}) at
assuming that all dipoles are parallel to the external electric field, and the inter-particle interaction does not change the direction of dipoles, we consider wave perturbations around the equilibrium state described with the following parameters $n=n_{0}$, $\textbf{v}=0$, $\textbf{E}=\textbf{E}_{ext}$. In the linear approximation on small simple harmonic perturbations $\delta f=f(\omega, \textbf{k})\exp(-\imath\omega t+\imath \textbf{k} \textbf{r})$ of the particle concentration $\delta n$, the velocity field $\delta \textbf{v}$, and the electric field $\delta \textbf{E}$, where $\omega$ is the frequency of the wave oscillation and $\textbf{k}$ is the wave vector.
Amplitudes of the wave perturbation satisfy to an algebraic form of the quantum hydrodynamic equations: the continuity equation (\ref{NI GP cont eq from GP})
\begin{equation}\label{NI GP cont eq from GP in set alg}
-\imath\omega \delta n(\omega, \textbf{k})+n_{0}\imath \textbf{k}\delta\textbf{v}(\omega, \textbf{k})=0,\end{equation}
the momentum balance equation (\ref{NI GP bal imp eq short from GP}) (the Euler equation)
$$-\imath\omega mn_{0}\delta v^{\alpha}(\omega, \textbf{k})+\frac{\hbar^{2}}{4m}\imath k^{\alpha}k^{2} \delta n(\omega, \textbf{k})$$
\begin{equation}\label{NI GP bal imp eq short from GP in set alg}=-gn_{0}\imath k^{\alpha}\delta n(\omega, \textbf{k})+
\imath n_{0}d k^{\alpha}\textbf{l}\delta\textbf{E}(\omega, \textbf{k}),\end{equation}
and the Maxwell equations (\ref{NI GP field good div}) and (\ref{NI GP field good curl})
\begin{equation}\label{NI GP field good div in set alg}k_{x}E_{x}(\omega, \textbf{k})+k_{y}E_{y}(\omega, \textbf{k})+k_{z}E_{z}(\omega, \textbf{k})=-4\pi d k_{z} \delta n(\omega, \textbf{k}),\end{equation}
and
\begin{equation}\label{NI GP field good curl in set alg}\textbf{k}\times\delta\textbf{E}(\omega, \textbf{k})=0.\end{equation}

The continuity equation (\ref{NI GP cont eq from GP in set alg}) gives a connection of the particle concentration amplitude $\delta n (\omega, \textbf{k})$ with the amplitude of the velocity field $\delta \textbf{v} (\omega, \textbf{k})$:
$$\delta n(\omega, \textbf{k})=n_{0}\frac{\textbf{k}\delta \textbf{v}(\omega, \textbf{k})}{\omega}.$$
Using this relation we find connection of the velocity field amplitude with the amplitude of the electric field from the Euler equation (\ref{NI GP bal imp eq short from GP in set alg}), which is
\begin{equation}\label{NI GP}(\textbf{k}\delta \textbf{v}(\omega, \textbf{k}))=\frac{-\omega d k^{2} \delta E_{z}(\omega, \textbf{k})}{m\omega^{2}-\frac{\hbar^{2}k^{4}}{4m}-gn_{0}k^{2}}.\end{equation}
We will put this formula in equation (\ref{NI GP field good div in set alg}), but now we express $\delta E_{x}$ and $\delta E_{y}$ via $\delta E_{z}$ using equation (\ref{NI GP field good curl in set alg}). We have
\begin{equation}\label{NI GP Ex-Ez}\delta E_{x}=\frac{k_{x}}{k_{z}}\delta E_{z},\end{equation}
and
\begin{equation}\label{NI GP Ey-Ez}\delta E_{y}=\frac{k_{y}}{k_{z}}\delta E_{z}.\end{equation}
These formulas we also put in the formula (\ref{NI GP field good div in set alg}). The formulas (\ref{NI GP Ex-Ez}) and (\ref{NI GP Ey-Ez}) are uncorrect if $k_{z}=0$. If $k_{z}=0$ we have $E_{z}=0$. Using it we find the Bogoliubov spectrum because electric dipoles give no contribution in the Euler equation, and consequently, in the particle evolution.

In result the equation (\ref{NI GP field good div in set alg}) appears in the following form
\begin{equation}\label{NI GP}k^{2}\delta E_{z}=\frac{4\pi d^{2}k_{z}^{2}k^{2}n_{0}\delta E_{z}}{m\omega^{2}-\frac{\hbar^{2}k^{4}}{4m}-gn_{0}k^{2}},\end{equation}
where $k^{2}=k^{2}_{x}+k^{2}_{y}+k^{2}_{z}$.

Assuming that the amplitude of the electric field is not zero we cancel $\delta E_{z}$ and get the dispersion equation
\begin{equation}\label{NI GP}\omega^{2}=\frac{\hbar^{2}k^{4}}{4m^{2}}+\frac{gn_{0}k^{2}}{m}+\frac{4\pi n_{0}d^{2}k^{2}_{z}}{m}.\end{equation}
This formula gives the dispersion of collective excitations in the electrically polarized three-dimensional BEC at assumption that all dipoles are parallel to the external electric field and the dipole-dipole interaction causes no evolution of the electric dipole direction (the polarization waves is the example of the coherent evolution of the dipole direction \cite{Andreev LongetDipFermi12}, \cite{Andreev TransDipBEC12}, \cite{Andreev PRB 11}).

\subsection{properties of the excitation dispersion}

Our calculation gives the following spectrum of the collective excitations
\begin{equation}\label{NI GP disp for discuss}\omega^{2}=\frac{\hbar^{2}k^{4}}{4m^{2}}+\frac{gn_{0}k^{2}}{m}+\frac{4\pi n_{0}d^{2}k^{2}\cos^{2}\theta}{m},\end{equation}
where $\cos\theta=k_{z}/k$.

This result differs from formula
\begin{equation}\label{NI GP disp from Lit}\omega^{2}=k^{2}\Biggl(\frac{n_{0}}{m}\biggl(g+\frac{C_{dd}}{3}(3\cos^{2}\theta-1)\biggr)+\frac{\hbar^{2}k^{2}}{4m^{2}}\Biggr)\end{equation}
obtained in the previous papers (see, for example, formula (5.1) in Ref. \cite{Lahaye RPP 09} or formula (11) in Ref. \cite{Baranov CR 12}), $C_{dd}$ is the dipolar coupling
constant $C_{dd}=d^{2}/\varepsilon_{0}$ in the SI units, where $\varepsilon_{0}$ is the vacuum
permittivity, or $C_{dd}=4\pi d^{2}$ in the CGS units. In formula (\ref{NI GP disp from Lit}) the constant of the short-range interaction depends on the electric dipole moment of molecule $g=g(d)$ \cite{Baranov CR 12} (see formula (3) and the text after it). To get correct properties of the spectrum we need to account this dependence explicitly (we present this consideration in the Appendix), but these properties have been considered with no account of dependence $g(d)$. We present the brief discussion of formula (\ref{NI GP disp from Lit}). For unpolarized BEC $g>0$ corresponds to the repulsive short-range interaction. The attractive short-range interaction $g<0$ leads to the unstable spectrum of unpolarized three-dimensional BEC. Let us discuss changes of the dispersion in the repulsive BEC $g>0$ caused by dipoles. The contribution of dipoles in the dispersion dependence may have different sign depending on $\theta$. As follows from formula (\ref{NI GP disp from Lit}) square of frequency become negative at large enough $C_{dd}$ and some angles, that reveals an instability of the three-dimensional polarized BEC. Under proper conditions the spectrum (\ref{NI GP disp from Lit}) reveals  a minimum in the dispersion at intermediate
momentum \cite{Santos PRL 03}. The formula (\ref{NI GP disp from Lit}) is obtained using the potential of dipole-dipole interaction in the absence of $\delta$-function term.

Let us discuss the spectrum (\ref{NI GP disp for discuss}) obtained in this paper.
At small wave vectors $k$ the frequency can be written as $\omega=\sqrt{(n_{0}/m)(g+4\pi d^{2} \cos^{2}\theta)}k$. We can see that, at the repulsive short-range interaction $g>0$, the coefficient $g+4\pi d^{2} \cos^{2}\theta$ is positive for all angles $\theta$. If $4\pi d^{2}/g >1$, at small angles $\theta\ll\pi/2$ the quantity $g+4\pi d^{2} \cos^{2}\theta$ is positive even for the repulsive short-range interaction $g<0$. It shows us that the electric polarization of BEC does not lead to destabilization of the spectrum of BEC.

\section{Bright soliton in the electrically polarized BEC}

The fundamental nonlinear excitation in attractive ($g<0$) unpolarized BEC is the bright soliton, which is an area of the increased concentration. We consider a perturbation of the hydrodynamic variables $n$, $\textbf{v}$, and $\textbf{E}$ as functions of $x-v_{0}t$ and assume that external electric field applied along $z$ axes. Therefore we consider propagation of the bright soliton perpendicular to the external field. We find that electric field gives no contribution in evolution of particle concentration $n$ and velocity field $\textbf{v}$. The last term in equation (\ref{NI GP bal imp eq short from GP}) is equal to zero, since $\partial_{x}E_{z}=0$ that follows from equation (\ref{NI GP field good div}). Consequently, we have no change in properties of the bright soliton.

Next, we consider the bright soliton propagation parallel to the external field, so we suppose that $n$, $\textbf{v}$, and $\textbf{E}$ are functions of $\xi=z-v_{0}t$. In this case the last term in the equation (\ref{NI GP bal imp eq short from GP}) appears as $nd\partial_{z}E_{z}=-4\pi d^{2}n\partial_{z}n$ that we find using equation (\ref{NI GP field good div}). Thus we can combine two terms in the right-hand side of the equation (\ref{NI GP bal imp eq short from GP}), which present $z$ projection of the force density $\textbf{F}$, and write it in the form of $F_{z}=-(g+4\pi d^{2})n\partial_{z}n$. We have that the dipole-dipole interaction of fully polarized electric dipoles leads to change the interaction constant. All difference, we find in compare with unpolarized case, is replacement of $g$ with $\tilde{g}=g+4\pi d^{2}$. Using the methods described in Ref.s \cite{Fedele EPJ.B.02}, \cite{Andreev Izv.Vuzov. 10} we come to the particle concentration profile in well-known form of
\begin{equation}\label{soliton solution in 1 order} n(\xi)=n_{0}\frac{1}{\cosh^{2}\alpha},\end{equation}
where $n_{0}=2m\mid E\mid/(\mid \tilde{g}\mid)$, $\alpha=\sqrt{2\mid E\mid}m\xi/\hbar$, where $E=c_{0}-v_{0}^{2}/2$ and $v_{0}$ is the velocity of the soliton propagation, $c_{0}$ is a constant. The bright soliton exists at $E<0$ and $\tilde{g}<0$.

The bright soliton exist at the condition $g+4\pi d^{2}<0$, which shows that the polarization leads to decreasing of the area of the bright soliton existing. In the electrically polarized BEC module of the short-range interaction constant must be more than $4\pi d^{2}$.

\section{Conclusion}

Explicitly considering the correct potential energy of the electric dipole-dipole interaction and assuming that all dipoles are parallel to the external field we presented a non-integral form of the GP equation for electrically polarized molecules and corresponding hydrodynamic equations (the continuity equation and the Euler equation). Solving these equations we obtained spectrum of collective excitations in the electrically polarized BEC. We found that the obtained spectrum differs from the one presented in literature. 3D uniform dipolar BEC does not reveal any instability for all directions of wave propagation. Introduction of the internal electric field satisfying the Maxwell equations allows to represent the GP equation in a nonintegral form and gives possibility to study the bright soliton analytically. The bright soliton in the electrically polarized BEC was considered using described approximation. We shown that area of parameters of the bright soliton existing is decreased in compare with the unpolarized BEC, but its form does not change, for the soliton propagation along the external field. And the polarization gives no influence on the bright soliton properties at the soliton propagation perpendicular to the external field.


\section{Appendix}
Writing "Schrodinger equation" for the condensate wave function, which appears as a part of $\psi$ operator \cite{Landau 9}, we have
$$\imath\hbar\partial_{t}\Phi(\textbf{r},t)=\Biggl(-\frac{\hbar^{2}}{2m}\triangle$$
\begin{equation}\label{NI GP A}+V_{ext}(\textbf{r})+\int\Phi^{*}(\textbf{r}',t)U(\textbf{r}-\textbf{r}')\Phi(\textbf{r}',t)d\textbf{r}'\Biggr)\Phi(\textbf{r},t)\end{equation}
Particles interact by means a short-range interaction and the long range dipole-dipole interaction, then
\begin{equation}\label{NI GP A}U=g\delta(\textbf{r}-\textbf{r}')+\frac{4\pi}{3}d^{2}\delta(\textbf{r}-\textbf{r}')+U_{dd0},\end{equation}
where $g=4\pi\hbar^{2}a/m$, $a$ is the scattering length describing short-range interaction existing at $d=0$, and
\begin{equation}\label{NI GP A}U_{dd0}=\frac{\textbf{d}\textbf{d}-3(\textbf{r}\textbf{d})(\textbf{r}\textbf{d})}{r^{3}}.\end{equation}
Delta functional terms can be combined $U=g'\delta(\textbf{r}-\textbf{r}')+U_{dd0}$, where $g'=g'(d)=g+4\pi d^{2}/3$. If amplitude of dipole-dipole interaction proportional to $d^{2}$ is large in comparison with the short-range interaction $g\ll d^{2}$, when we can write $g'=4\pi d^{2}/3$. Using explicit form of $g'(d)$ and $C_{dd}$ we rewrite formula (\ref{NI GP disp from Lit}) and find
\begin{equation}\label{NI GP A}\omega^{2}=\frac{n_{0}k^{2}}{m}\biggl(g+\frac{4\pi}{3}d^{2}+\frac{4\pi d^{2}}{3}(3\cos^{2}\theta-1)\biggr)+\frac{\hbar^{2}k^{4}}{4m^{2}},\end{equation}
combining together terms proportional to $d^{2}$ we get formula (\ref{NI GP disp for discuss}) obtained in this paper.

\begin{acknowledgements}
The author wish to thank Professor L.S. Kuz'menkov for discussion of the results obtained.
\end{acknowledgements}


\begin{thebibliography}{17}


\bibitem{Yi PRA 00} S. Yi and L. You, Phys. Rev. A, \textbf{61}, 041604(R) (2000).
\bibitem{Goral PRA 00} K. Goral, K. Rzazewski, and T.
Pfau, Phys. Rev. A \textbf{61}, 051601(R) (2000).
\bibitem{Santos PRL 00} L. Santos, G.V. Shlyapnikov, P. Zoller, and M.
Lewenstein, Phys. Rev. Lett. \textbf{85}, 1791 (2000).


\bibitem{Yi PRA 01} S. Yi and L. You, Phys. Rev. A, \textbf{63}, 053607 (2001).
\bibitem{Goral PRA 02} K. Goral and L. Santos, Phys. Rev. A \textbf{66}, 023613
(2002).


\bibitem{Santos PRL 03} L. Santos, G. V. Shlyapnikov, and M. Lewenstein, Phys. Rev.
Lett. \textbf{90}, 250403 (2003).
\bibitem{Dell PRL 03} D. H. J. O'Dell, S. Giovanazzi, and G. Kurizki, Phys. Rev.
Lett. \textbf{90}, 110402 (2003).
\bibitem{Giovanazzi EPJD 04} S. Giovanazzi and D. H. J. O'Dell, Eur. Phys. J. D \textbf{31}, 439
(2004).

\bibitem{Ronen PRL 07} S. Ronen, D. C. E. Bortolotti, and J. L. Bohn, Phys. Rev.
Lett. \textbf{98}, 030406 (2007).
\bibitem{Jona-Lasinio 13} M. Jona-Lasinio, K.  Lakomy and L. Santos,  arXiv:1301.4907.




\bibitem{Fischer PRA 06R} Uwe R.
Fischer, Phys. Rev. A \textbf{73}, 031602(R) (2006).



\bibitem{Lahaye RPP 09} T. Lahaye, C. Menotti, L. Santos, M. Lewenstein, and T. Pfau, Rep. Prog. Phys. \textbf{72}, 126401 (2009).

\bibitem{Wilson PRL 08} R.M. Wilson, S. Ronen, J.L. Bohn, and H. Pu, Phys.
Rev. Lett. \textbf{100}, 245302 (2008).

\bibitem{Ticknor PRL 11} C. Ticknor, R.M. Wilson, and J.L. Bohn, Phys. Rev. Lett. \textbf{106}, 065301 (2011).

\bibitem{Wilson PRA 12} R.M. Wilson, C. Ticknor, J.L. Bohn, and E. Timmermans,
Phys. Rev. A \textbf{86}, 033606 (2012).


\bibitem{Baranov CR 12} M. A. Baranov, M. Dalmonte, G. Pupillo, and P. Zoller, Chem. Rev.,  \textbf{112},  5012 (2012). 
\bibitem{Carr NJP 09} L. D. Carr et al., New J. Phys. \textbf{11}, 055049 (2009).
\bibitem{Nozieres JLTP 06-09} P. Nozieres, J. Low Temp. Phys. \textbf{142}, 91 (2006); ibid
\textbf{156}, 9 (2009).

\bibitem{Sieberer ar 11} L.M. Sieberer, and M.A.Baranov, arXiv:1110.3679.





\bibitem{Andreev arxiv 12 02} P. A. Andreev and L. S. Kuz'menkov, arXiv:1201.2440.
\bibitem{Andreev arxiv Pol 11} P. A.  Andreev and L. S. Kuzmenkov, arXiv:1106.0822.
\bibitem{Andreev RPJ 12} P. A. Andreev, Russian Physics Journal \textbf{54}, 1360 (2012).

\bibitem{Andreev LongetDipFermi12} P. A. Andreev, arXiv:1209.4196.
\bibitem{Andreev TransDipBEC12} P. A. Andreev, L. S. Kuz'menkov, arXiv:1208.1000.

\bibitem{Andreev PRB 11} P. A. Andreev, L. S. Kuz'menkov, M. I. Trukhanova, Phys. Rev. B \textbf{84}, 245401 (2011).




\bibitem{MaksimovTMP 1999} L. S. Kuz'menkov and S. G. Maksimov,  Teor. i Mat. Fiz.,
\textbf{118} 287 (1999) [Theoretical and Mathematical Physics \textbf{118} 227 (1999)].
\bibitem{MaksimovTMP 2001} L. S. Kuz'menkov, S. G. Maksimov, and V. V. Fedoseev, Theor.
Math. Fiz. \textbf{126} 136 (2001) [Theoretical and Mathematical
Physics, \textbf{126} 110 (2001)].
\bibitem{Andreev IJMP 12} P. A. Andreev, L. S. Kuz'menkov,  Int. J. Mod. Phys. B \textbf{26} 1250186 (2012).

\bibitem{Andreev PRA08} P. A. Andreev, L. S. Kuz'menkov, Phys. Rev. A \textbf{78}, 053624
(2008).
\bibitem{Andreev IJMP B 13} P. A. Andreev, Int. J. Mod. Phys. B \textbf{27}, 1350017 (2013).

\bibitem{Landau 2} L.D. Landau and E.M. Lifshitz, \textit{The Classical Theory of Fields} (Butterworth-Heinemann, 1975).

\bibitem{Landau 9} L.D. Landau and E.M. Lifshitz, \textit{Statistical Physics, Part 2} (Butterworth-Heinemann, 1980).

\bibitem{Fedele EPJ.B.02} R. Fedele, H. Schamel, Eur. Phys. J. B \textbf{27}, 313 (2002).

\bibitem{Andreev Izv.Vuzov. 10} P. A. Andreev, M. I. Trukhanova,
Russian Physics Journal \textbf{53}, 1196 (2011).



%
\end{thebibliography}
\end{document}